\documentclass[%
 reprint,
 superscriptaddress,preprintnumbers,
 nofootinbib,
 amssymb,
 aps,
]{revtex4-1}

\usepackage[utf8]{inputenc}
\usepackage[colorlinks=true,citecolor=blue,linkcolor=blue]{hyperref}
\usepackage[normalem]{ulem}
\usepackage{amsmath,amssymb,mathtools}
\usepackage{epsfig}
\usepackage{graphicx}               
\usepackage{url}
\usepackage{color,colortbl}
\usepackage{slashed}
\usepackage{cancel}
\usepackage{multirow}
\usepackage{placeins}
\usepackage[dvipsnames]{xcolor}
\usepackage{epstopdf}
\usepackage{soul}
\usepackage{tikz}
\usepackage{tikz-feynman}
\usepackage[capitalise, english]{cleveref}
\usepackage{siunitx}
\usepackage{xspace}
\usetikzlibrary{trees}
\usetikzlibrary{decorations.pathmorphing}
\usetikzlibrary{decorations.markings}

\definecolor{red}{rgb}{0.6,.0706,.1373}
\definecolor{blue}{rgb}{0,0.396,0.741}
\newcommand\myshade{80}
\colorlet{mylinkcolor}{violet}
\colorlet{mycitecolor}{violet}
\colorlet{myurlcolor}{violet}

\hypersetup{
  linkcolor  = mylinkcolor!\myshade!black,
  citecolor  = mycitecolor!\myshade!black,
  urlcolor   = myurlcolor!\myshade!black,
  colorlinks = true
}

\allowdisplaybreaks

\newcommand{\Z}{\mathbb{Z}}
\newcommand{\R}{\mathbb{R}}
\newcommand{\be}{\begin{equation}}
\newcommand{\ee}{\end{equation}}
\newcommand{\bea}{\begin{eqnarray}}
\newcommand{\eea}{\end{eqnarray}}
\newcommand\aNLO{{\sc\small MadGraph5\_aMC@NLO}}

\def\beq#1\eeq{\begin{align}#1\end{align}}

\makeatletter
\providecommand*{\diff}%
  {\@ifnextchar^{\DIfF}{\DIfF^{}}}
\def\DIfF^#1{%
  \mathop{\mathrm{\mathstrut d}}%
    \nolimits^{#1}\gobblespace}
\def\gobblespace{%
  \futurelet\diffarg\opspace}
\def\opspace{%
  \let\DiffSpace\!%
  \ifx\diffarg(%
    \let\DiffSpace\relax
  \else
    \ifx\diffarg[%
      \let\DiffSpace\relax
    \else
        \ifx\diffarg\{%
        \let\DiffSpace\relax
      \fi\fi\fi\DiffSpace}


\begin{document}

\title{Topological Portal to the Dark Sector}

\begin{flushright}
 CERN-TH-2024-010
\end{flushright}

\author{Joe Davighi}
\email{joseph.davighi@cern.ch}
\affiliation{Theoretical Physics Department, CERN, 1211 Geneva 23, Switzerland}
\author{Admir Greljo}
\email{admir.greljo@unibas.ch}
\affiliation{Department of Physics, University of Basel, Klingelbergstrasse 82, CH 4056 Basel, Switzerland}
\author{Nud\v{z}eim Selimovi\'c}
\email{nudzeim.selimovic@pd.infn.it}
\affiliation{Istituto Nazionale di Fisica Nucleare, Sezione di Padova, Via Francesco Marzolo 8, 35131 PD, Italy}


\preprint{}

\begin{abstract}
We propose a unique topological portal between quantum chromodynamics (QCD) and a dark sector characterized by a global symmetry breaking, which connects three QCD to two dark pions.
When gauged, it serves as the leading portal between the two sectors, providing an elegant, self-consistent scenario of light thermal inelastic dark matter. The inherent antisymmetrization leads to diminished annihilations at later times and suppressed direct detection. However, novel collider signatures offer tremendous prospects for discovery at Belle II.

\end{abstract}

\maketitle

\section{Introduction} \label{sec:intro}

The quest to unravel the mysteries of dark matter (DM) stands as a pivotal challenge in contemporary physics. 
That DM is composed of a hidden particle physics sector is a well-motivated hypothesis~\cite{ParticleDataGroup:2022pth}.
While DM has primarily revealed its presence through gravitational interactions, the existence of other \textit{portals} may not only be feasible but perhaps essential in elucidating aspects of cosmological evolution and accounting for the observed relic abundance~\cite{Gondolo:1990dk}. 

In this letter, we propose a novel portal that links quantum chromodynamics (QCD) and a dark sector characterized by a global symmetry breaking.
Our starting point is to seek a hitherto overlooked topological interaction between the two theories in their confined phases.
After a comprehensive and rigorous 
analysis of the possible global symmetries characterizing QCD-like dark sectors, we discover that there is a {\em unique} coset structure featuring a non-trivial Wess--Zumino--Witten (WZW) term~\cite{Wess:1971yu, Witten:1983tw} which connects three QCD pions to two dark pions.

Having established the existence and uniqueness of this topological portal operator, we delve into the resulting phenomenology. The topological portal enables dark number conservation through a $\mathbb{Z}_2$ symmetry, ensuring the stability of the lightest dark pion. Furthermore, after quantum electrodynamics (QED) is properly included, the portal operator yields 4-point interactions involving one photon, one neutral QCD pion ($\pi^0$ or $\eta$), and two dark pions, see Fig.~\ref{fig:feynman}.
The chiral Lagrangian power counting identifies these (appropriately gauged) WZW terms as the {\em leading} portal between the two sectors, which thus defines the dominant phenomenology. Prompted by this, we investigate whether such an operator can establish the correct relic abundance. Intriguingly, we uncover a self-consistent scenario whereby DM achieves thermal freeze-out at temperatures below the QCD phase transition. 
This phenomenon is primarily governed by late-stage interactions via the topological operator, hinting at the presence of DM in the GeV range~\cite{Zurek:2024qfm}.

{The topological nature of this novel DM portal is more than a theoretical nicety; it plays a crucial role in shaping a consistent phenomenology of light thermal dark matter.}
The topological operator is obtained by integrating a differential form, and is therefore completely antisymmetric under exchanging pairs of fields. Thus, it necessarily couples two {\em different} dark pion species. After freeze-out governed by $\chi_1 \chi_2$ coannihilations via the diagram on the right of Fig.~\ref{fig:feynman}, the heavier dark pion rapidly decays, leaving the lighter as the residual DM. This elegantly leads to diminished DM annihilation into SM particles at later times for the topological portal, skirting otherwise stringent annihilation constraints from CMB anisotropies on elastic $s$-wave scattering~\cite{Galli:2011rz, Planck:2018vyg}. In addition, direct detection experiments lack sensitivity due to the kinematic suppression of inelastic scatterings. Thus, we stumble upon an elegant and natural realization of the light thermal inelastic DM scenario~\cite{Tucker-Smith:2001myb, Izaguirre:2015zva}. 

The key to probing this new portal lies in collider experiments. Novel and previously unexplored collider signatures are anticipated in current \( e^+ e^- \) flavor factories, calling for the design of new search strategies. Depending on the mass splitting, the heavy dark pion exhibits decays that are either displaced or detector-stable. The future at Belle II looks incredibly bright, offering the potential to explore much of the intriguing parameter space defined by the relic abundance.

\section{Topological portal EFT}

\begin{figure}
    \centering
    \includegraphics[width=0.9\linewidth]{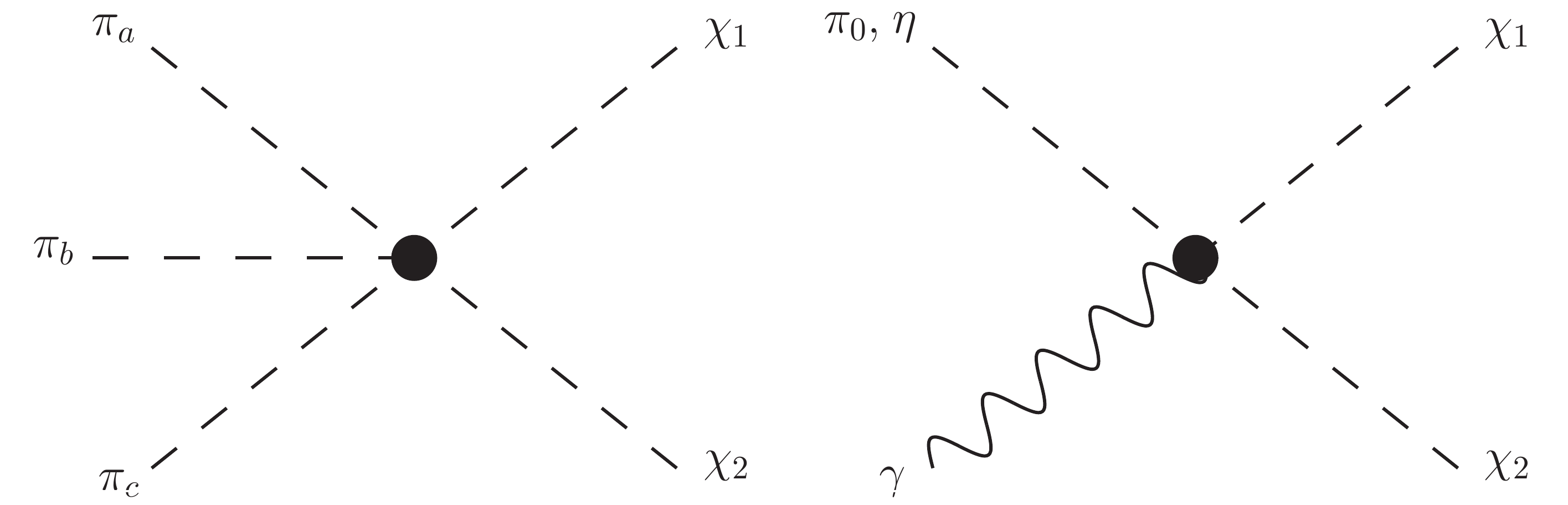}
   \caption{Feynman diagrams for the two scattering channels mediated by the topological portal, where $\pi_a$ denote QCD pions and $\chi_i$ denote dark pions on an $S^2$ coset space. The right-hand diagram mediates a $2\to 2$ co-annihilation that sets the observed dark matter relic abundance, with the lighter of the two dark pions eventually remaining as dark matter. The inherent antisymmetry of the topological interaction forbids corresponding interactions involving $\chi_1\chi_1$ or $\chi_2\chi_2$. }
    \label{fig:feynman}
\end{figure}

We consider a dark sector characterized by a global symmetry-breaking transition $K \to H$, delivering dark pions as the DM. The relevant degrees of freedom are the QCD pions plus dark pions, which are collectively described by a 4-d non-linear sigma model 
on a product coset  
\be \label{eq:coset}
X = \frac{SU(3)_L \times SU(3)_R \times K}{SU(3)_{L+R} \times H} \cong SU(3) \times \frac{K}{H}\, ,
\ee
with global symmetry $G=SU(3)_L \times SU(3)_R \times K$. In addition to the usual terms appearing in the effective field theory (EFT) construction of Callan, Coleman, Wess, and Zumino~\cite{Callan1969}, which require a $G$-invariant metric on $X$, 
the action also admits 
$G$-invariant {\em topological interactions} that can be constructed without a metric. {Such topological interactions can instead be obtained by directly integrating a {\em differential form}, which recall is a totally antisymmetric covariant tensor, on the coset space $X$.}
In this paper, we identify EFTs in which such a topological interaction provides a {qualitatively new kind of} portal through which DM interacts with the SM, indicated by the diagrams in Fig.~\ref{fig:feynman}.


Recall that in pure QCD there already appears a topological interaction, the WZW term~\cite{Wess:1971yu, Witten:1983tw}. 
This term originates from the existence of an $SU(3)_{L}\times SU(3)_{R}$-invariant 
differential 5-form $\omega_5=\frac{-i}{480\pi^3} N_c \mathrm{Tr}\left(g^{-1} dg \right)^5$ on $SU(3)$, where $g$  is the $SU(3)$-valued pion field {and $d$ is the exterior derivative}, {that is {\em closed}, meaning $d\omega_5=0$.} 
The action, evaluated on 4-d spacetime $\Sigma$, can be defined by extending $g(x)$ to a 5-d bulk manifold $X$ whose boundary is $\Sigma$ and integrating $\omega_5$ thereon. Requiring the path integral phase $e^{2\pi i\int_X \omega_5}$ be independent of the choice of the bulk $X$ forces the coefficient $N_c$ be an integer~\cite{Witten1983,Lee:2020ojw,Freed:2006mx}. 
Famously, $N_c$ is fixed to be the number of colors in QCD by anomaly matching.

In addition to the invariant closed 5-form, QCD also features an invariant closed 3-form, $\omega_3 \propto \mathrm{Tr} (g^{-1} dg)^3$. In fact, there are {\em no other $SU(3)_{L}\times SU(3)_{R}$  invariant forms} on $SU(3)$ with which one could construct topological terms involving QCD pions. Simply by virtue of its degree, $\omega_3$ does not appear in the pure QCD action. 
By coupling to {particular choices of dark coset such as $K/H\cong SU(2)/SO(2)\cong S^2$}, {the QCD-invariant 3-form} $\omega_3$ can be used to construct a second topological term {that connects QCD to the dark sector}. 
{That term, whose construction involves 
mathematical details which we relegate to the End Matter, is
\begin{align} \label{eq:action-new}
  &S[\Sigma = \partial X] = n \int_{X} \tilde\omega_3 \wedge \Omega_2\, , \quad \text{where} \\
  \tilde\omega_3 = &\frac{1}{24\pi^2} \left[\mathrm{Tr}\left(g^{-1} dg \right)^3 -6eF\wedge 
  \mathrm{Tr}\left(Q g^{-1}dg\right)\right] \, , \label{eq:3-form-new}
\end{align}
the matrix $Q=t_3+t_8/\sqrt{3}$ generates the electromagnetic subgroup $U(1)_Q \subset SU(3)_{L+R}$ of the QCD flavour symmetry,
and $\Omega_2$ is the volume form on the $K/H=S^2$ dark coset (normalized such that $\int_{S^2}\Omega_2=1$).
The coefficient $n\in \Z$ must be quantized for this to define a consistent low-energy effective action {\em i.e.} to be independent of the choice of bulk extension $X$, like the coefficient of the WZW term in QCD. This $n$ will be determined by the UV completion, which we do not specify in this work.
}

{The action~\eqref{eq:action-new} contains two terms, coming from the two pieces in~\eqref{eq:3-form-new}. The first term can be written purely using the QCD pion fields, whereas the second term is the result of turning on the gauge field $F=dA$ for QED, and so gives rise to a new interaction connecting the photon to the dark matter.}

{Let us now unpack the physics of these interactions. To do so, we expand the 5-form $\tilde\omega_3\wedge \Omega_2$ in terms of the pion and dark pion fields. }
First, consider the QCD part.
Expanding $g=\exp(2i \pi_a(x) t_a/f_\pi)$ where $t_a=\frac{1}{2}\lambda_a$ with $\lambda_a$ being the Gell-Mann matrices, $g^{-1}dg = \frac{2i}{f_\pi}t_a d\pi_a + \mathcal{O}(\pi^2)$ implies 
\begin{align}
    \mathrm{Tr}(g^{-1} dg)^3 
    &= (2/f_\pi^3)f_{abc}\, d\pi_a\wedge  d\pi_b\wedge  d\pi_c + \mathcal{O}(\pi^4)\, ,
\end{align}
where $f_{abc}$ are
the $SU(3)$ structure constants defined such that $[t_a,t_b]=if_{abc}t_c$, and where we use antisymmetry of the wedge product to write $\mathrm{Tr}(t_a t_b t_c)\mapsto \mathrm{Tr}(t_a [t_b, t_c])/2$, which is proportional to $f_{abc}$. Our convention is such that the QCD pion kinetic term is $f_\pi^2\mathrm{Tr}(\partial_\mu g\, \partial^\mu g^{-1})/4$, with $f_\pi = 92$ MeV.
Now consider the `dark part'. Taking $f_D$ to be the dark pion decay constant, let $\chi_1/f_D$ and $\chi_2/f_D$ be local coordinates on $S^2$ corresponding to canonically normalised fields in the vicinity of the vacuum point ($\chi_i=0$). 
Then $\Omega_2 = \cos(\chi_1/f_D) d\chi_1 \wedge d\chi_2/(4\pi f_D^2) = d\chi_1 \wedge d\chi_2/(4\pi f_D^2) + \mathcal{O}(\chi^3)$.  

We can substitute these expressions into~\eqref{eq:action-new}. Because the 5-forms we integrate are closed, they can be locally written as the derivative of a 4-form potential (by the Poincar\'e lemma), allowing us to write a local expression for the action as a 4-d integral, as follows:
\begin{align} \label{eq:action-new-2}
    S[\Sigma] = \int_\Sigma &\left[\frac{in \epsilon^{\mu\nu\rho\sigma}}{48\pi^2 f_\pi^3 f_D^2}  f_{abc} \epsilon_{ij} \pi_a \partial_\mu \pi_b \partial_\nu \pi_c \partial_\rho \chi_i \partial_\sigma \chi_j \right. \\
    &+\left.\frac{ne\, \epsilon^{\mu\nu\rho\sigma}}{16\pi^2 f_\pi f_D^2}\left(\pi^0 + \frac{\eta}{\sqrt{3}}\right)F_{\mu\nu} \partial_\rho\chi_1  \partial_\sigma\chi_2 \right]\,. \nonumber
\end{align}
These two topological interactions, which together constitute the `topological portal', are depicted in Fig.~\ref{fig:feynman}.

We remark that very little would be different if we take the QCD coset to be $SU(2)$ instead of $SU(3)$, {\em i.e.} considering only the pion triplet: $f_{abc}$ are simply replaced by $\epsilon_{ijk}$ in \eqref{eq:action-new-2}. 
In fact, the phenomenology will be driven by $SU(2)$ isospin, which is a better approximate symmetry of QCD.


As an aside, it is
helpful to contrast the topological portal~\eqref{eq:action-new} with a different topological interaction considered in strongly-interacting massive particle (or `SIMP') dark matter~\cite{Hochberg:2014dra,Hochberg:2014kqa}. The SIMP scenario invokes a WZW term purely in the dark sector, mirroring the one in QCD, to mediate $3\chi\to2\chi$ processes; the communication with the SM then proceeds via a {\em non}-topological interaction. In our scenario, a `mixed' topological term involving both SM and dark pions, which has no analog in QCD itself, directly provides the portal to the dark sector.

The second term in~\eqref{eq:action-new-2}, which couples $\gamma \pi^0 (\eta)$ to the pair of dark pions, is an effective dimension-7 operator in our EFT. 
Regarding EFT power-counting, this is the lowest-dimension interaction involving both QCD and dark pions. The next such interaction occurs at dimension-8, namely $\frac{1}{f_\pi^2 f_D^2}(D_\mu \pi_a D^\mu \pi^a) (\partial_\nu \chi_i \partial^\nu \chi^i)$, where $D_\mu=\partial_\mu + ieA_\mu$ is the covariant derivative.
If the relevant energy scales are low with respect to $f_\pi$ and $f_D$, one might expect the topological portal to be the leading portal in this EFT.

We remark that the existence of this topological portal places conditions on the dark coset $K/H$, as we derive in the End Matter. The upshot is that $K/H$ must have non-vanishing 2$^{\text{nd}}$ de Rham cohomology, which is, in fact, rather uncommon for coset spaces used in dark sector model building 
such as $SU(n)$, $SU(2n)/Sp(2n)$, or $SU(n)/SO(n)$ for $n>2$. This makes the topological portal a highly non-generic and, hence, all the more intriguing possibility that has been overlooked in all dark sector theories to date.



%

\section{Relic abundance from the topological portal}
\label{sec:relic}

We are now ready to investigate the phenomenology of the topological portal in cosmology and at colliders. For simplicity, 
we study the Boltzmann equations~\cite{Gondolo:1990dk} in the approximation of negligible mass splitting, commenting on the mass splitting at the end of this section. We describe the evolution of $\chi_1+\chi_2$ yield, $Y_\chi$, defined as the ratio of the combined dark pion number density $n_\chi$ to the entropy density~$s$,
\begin{equation}
\frac{dY_{\chi}}{dx} = - \sqrt{\frac{\pi g_{*}}{45}} \frac{M_{\rm P}m_\chi}{x^2}\langle \sigma v
\rangle \left(Y^2_\chi-Y_{\rm eq}^2\right)\,.
\label{eq:Boltzmann}
	\end{equation}
The evolution parameter is $x = m_\chi/T$ where $m_\chi$ is the dark pion mass, $M_{\rm P}$ is the Planck mass, and $g_{*}$ is the effective number of relativistic degrees of freedom. The thermally averaged cross section times the M{\o}ller velocity, $v
=\sqrt{|\mathbf{v}_1-\mathbf{v}_2|^2 - |\mathbf{v}_1\times\mathbf{v}_2|^2}$, can be expressed as a single integral~\cite{Gondolo:1990dk}
\begin{equation}
		\langle \sigma v
		\rangle = \frac{\int_{4m_\chi^2}^{\infty}\sigma\sqrt{s}(s-4m_\chi^2)\,{\rm{K}}_1(\sqrt{s}/T)\,{\rm{d}}s}{8m_\chi^4 T\,{\rm K}_2^2(m_\chi/T)}\,,
		\label{eq:thavgxsec}
\end{equation}
with $\sigma$ being the unpolarised cross section for the process $\chi_1\chi_2\to\pi^0\gamma$, $s=(p_{\chi_1}+p_{\chi_2})^2$, and ${\rm K}_i$ are the modified Bessel functions of the $i$-th order. Finally, the equilibrium yield $Y_{\rm eq}$ is a function of the evolution parameter $x$, namely $Y_{\rm eq} = {45 x^2 {\rm K}_2(x)}/(4\pi^4 g_*)$.

For our topological portal, the  unpolarised cross section is
		$\sigma = \frac{n^2}{1536 \,\pi^4}\, \frac{\alpha_{\rm Q}}{f_\pi^2 f_D^4}\,s^{3/2}\sqrt{s-4m_\chi^2}$,
where $\alpha_{\rm Q} = e^2/(4\pi)$ is the QED coupling constant, and we summed over the photon polarisations. After the integration in Eq.~\eqref{eq:thavgxsec}, we obtain
\begin{equation}
		\langle \sigma v\rangle = \frac{n^2}{64 \,\pi^{7/2}}\,\frac{\alpha_{\rm Q}x_f^4}{f_\pi^2}\, \frac{x\, {\rm G}^{3,0}_{1,3}\left(\genfrac{}{}{0pt}{}{-2}{-\frac{9}{2}\,-\frac{1}{2}\,\frac{1}{2}}|\,x^2\right)}{{\rm K}_2^2(x)}\,,
		\label{eq:sigmav}
\end{equation}
where $x_f=m_\chi/f_D$, and ${\rm G}(x^2)$ is the Meijer G-function~\cite{ADAMCHIK1995283}.

\begin{figure}[t]
	\centering
	\includegraphics[width=0.46\textwidth]{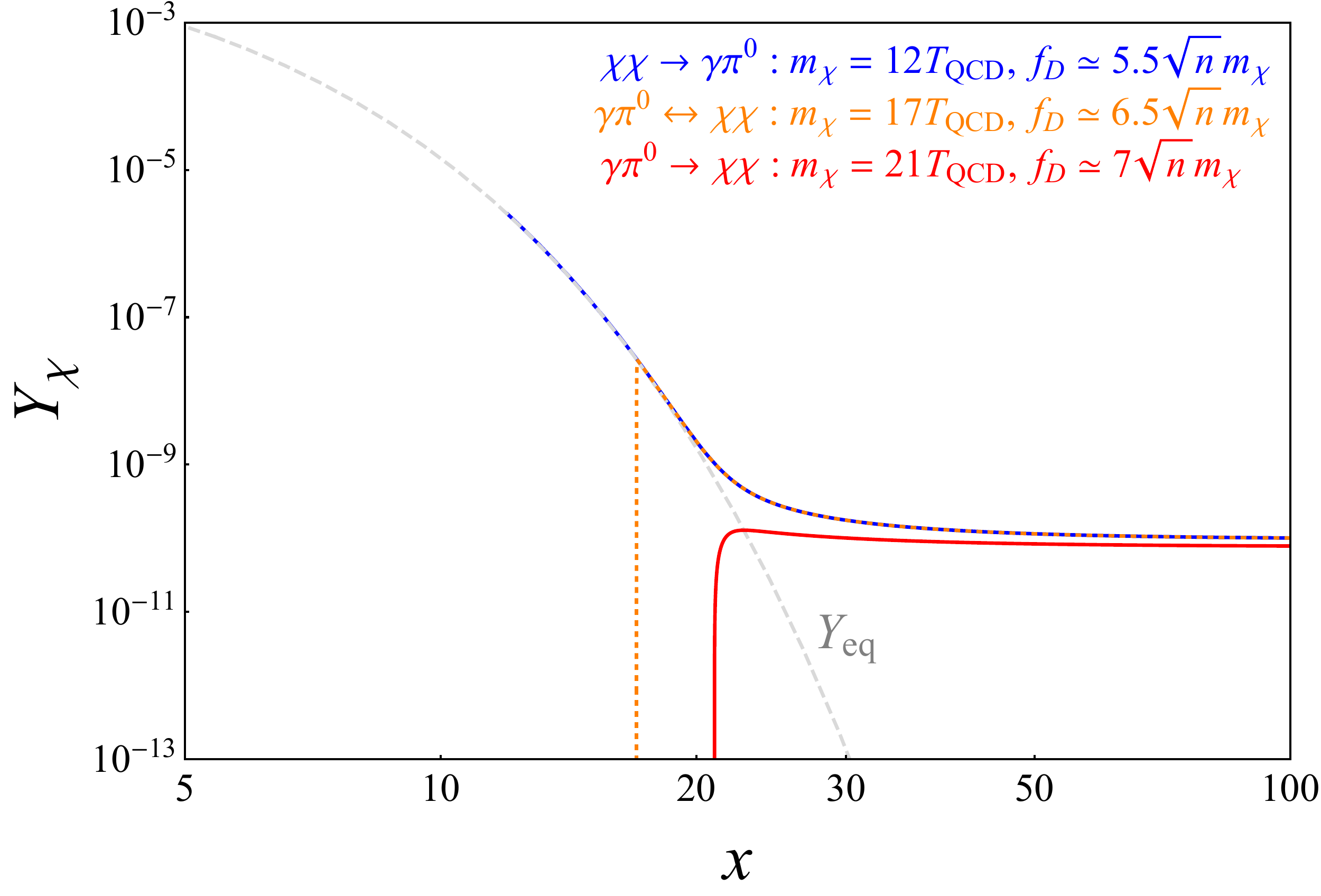}
	\caption{The dark pion yield as a function of $x=m_\chi/T$. The different lines show qualitatively different possibilities for setting the relic abundance, depending on the cosmological history of dark pions. See the main text for details.}
	\label{fig:relics}
\end{figure}
 
Depending on the cosmological history of the dark pions, Eq.~\eqref{eq:Boltzmann} can describe different scenarios setting the dark pion relic abundance. On the one hand, if the dark pions establish thermal equilibrium with the bath before the topological operator turns on at the QCD phase transition, then standard DM freeze-out occurs via $\chi\chi\to\gamma\pi^0$ at $x\sim 23$, after the topological portal is activated. This is shown by the blue line in Fig.~\ref{fig:relics}, for which the topological operator turns on at $x_{\rm QCD} :=m_\chi/T_{\rm QCD} = 12$. 
On the other hand, if the initial yield of the dark pions is negligible, and they are produced only after the QCD phase transition, they proceed with a quick thermalization followed by the freeze-out, and the correct relic abundance can still be attained. This is shown by an orange line in  Fig.~\ref{fig:relics}, for which the QCD phase transition occurs at $x=x_{\rm QCD}=17$.\\ 
\indent Consequently, the correct yield today ($Y_{\chi}$ for $x \to\infty$) can be achieved {\em irrespective} of the dark pion cosmological history. The topological portal thus robustly sets the correct relic abundance,
\begin{equation}
		\Omega_\chi h^2 \approx \frac{2\, x_{\rm QCD}\,T_{\rm QCD}\, Y_{\infty}}{3.6\cdot 10^{-9} \,{\rm GeV}} \approx 0.12\,,
\end{equation}
through an interplay between the QCD phase transition onset relative to the dark pion mass, determined by $x_{\rm QCD}$, and the $x_f=m_\chi/f_D$ ratio. In both scenarios, the QCD phase transition should happen no later than $x=x_{\rm max}=23$; after this point, the dark pion yield drops below the value necessary to account for the DM relic abundance. This limiting case is shown by a red line in Fig.~\ref{fig:relics}. The upper bound on $x_{\rm QCD}$ translates to an upper bound on $m_\chi$. Specifically, to explain the relic abundance via the topological portal points towards \textit{light thermal DM} with $m_\chi \lesssim 3.7$ GeV. \\
\indent In the computation, we used the QCD phase transition temperature, $T_{\rm QCD}\sim 160\,\, {\rm MeV}$, and the number of effective degrees of freedom, $g_*\sim 18$, assuming the topological operator turns on only after the QCD phase transition~\cite{ParticleDataGroup:2022pth}. While larger freeze-out temperatures are not inherently problematic, a correct description in this context would require full QCD rather than chiral perturbation theory. The fact that the correct relic abundance demands $f_D / m_\chi \sim \mathcal{O}(5.5- 7)\times \sqrt{n}$ is consistent with the dark pions being the lightest dark states. 

\begin{table}[t]
  \centering
  \begin{tabular}{|c|c|c|}
    \hline
    $\Delta m_\chi$ & $\lesssim 1.7 m_{\pi^0}$ & $\,\,\gtrsim 1.7 m_{\pi^0} \,\,\,$ \\
    \hline
    Signature & $\pi^0 +\slashed{E}_T$ & $\pi^0+\slashed{E}_T+\text{DV}(\pi^0\gamma \slashed{E}_T)$\\
    \hline
  \end{tabular}
  \caption{Collider signatures. Here `DV' indicates a displaced secondary vertex. The values of $\Delta m_\chi$ for which $\chi_2$ lifetime is approximately $10^{-7}$\,sec depend on the value of $m_{\chi_1}$, and vary from $[1.3 - 2.1] \,m_{\pi^0}$ in the mass region $m_{\chi_1} \in [1 - 3.5]\,$GeV.} 
  \label{tab:Collider_sig} 
\end{table}

The non-zero mass splitting, $\Delta m_\chi = m_{\chi_2}- m_{\chi_1}$, leads to a suppression of the co-annihilation cross section. The dominant effect can be captured by introducing an exponential suppression factor $e^{- x \Delta }$ in Eq.~\eqref{eq:sigmav}, where $\Delta := \frac{\Delta m_{\chi}}{ m_{\chi_1}}$~\cite{Griest:1990kh, DAgnolo:2018wcn}. The resulting $f_D$ which fits the relic abundance is, accounting for the mass splitting,
\begin{equation}\label{eq:coannn}
    f_D(\Delta) \approx f_D(0) \, e^{- \frac{ x_{\rm max} \Delta} {4} }~,
\end{equation}
where the factor of $1/4$ appearing in the exponent comes from the dependence $\langle \sigma v \rangle \propto x_f^4$ in Eq.~\eqref{eq:sigmav}.
This is a numerically large suppression effect that we account for in understanding the viable parameter space;
for example, for $\Delta = 5\times \frac{m_{\pi^0}}{3.5\,\text{GeV}}$ one finds $f_D(\Delta)/f_D(0) \approx 1/3$.

\section{Direct and indirect detection}

A central feature of the topological portal scenario is that, being a differential form, it allows only for totally {\em antisymmetric} interactions between the two dark pions.  
With a non-zero $\Delta m_\chi$, $\chi_2$ decays to $\chi_1$ shortly after freeze-out, leaving $\chi_1$ as the sole dark matter component. Then, given the antisymmetry property, DM annihilations at later times are suppressed because they require an interaction of $\chi_1$ with $\chi_2$ at leading order; $\chi_1 \chi_1 \to $ SM  occurs only at higher-order, and thus easily avoids otherwise stringent indirect detection constraints from CMB anisotropies~\cite{Galli:2011rz, Planck:2018vyg}. 
Likewise, direct detection experiments are ineffective for the inelastic $\chi_1 \to \chi_2$ up-scattering, {given $\Delta m_\chi = \mathcal{O}(\text{GeV})$ while the local dark matter velocity is non-relativistic.} 
As explained, elastic scattering is a higher-order process with an extremely small cross-section.

\section{Novel collider phenomenology}

\begin{figure}
    \centering
    \includegraphics[width=0.7\linewidth]{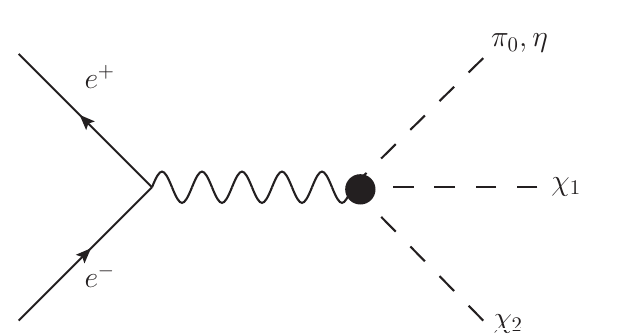}
    \caption{Production of dark matter via the topological portal, indicated by the blob as in Fig.~\ref{fig:feynman}), at a low-energy $e^+e^-$ collider experiment like Belle II.}
    \label{fig:feyn-BelleII}
\end{figure}

How, then, can we test this scenario? Interestingly, collider experiments offer a promising avenue through $e^+ e^- \to \gamma^* \to \chi_1 \chi_2 \pi^0$ production, see Fig.~\ref{fig:feyn-BelleII}. In particular, the required collider energy, high luminosity, and hermetic environment of Belle II result in exceptional sensitivity, potentially covering the full parameter space set by the DM relic abundance. Different mass splittings $\Delta m_\chi$ suggest different search strategies, as summarised in Tab.~\ref{tab:Collider_sig}. For a small mass splitting, $\chi_2$ is detector stable resulting in the final state with $\pi^0$ and missing energy $\slashed{E}_T$; otherwise, $\chi_2$ decays to $\pi^0$, photon, and missing energy at a displaced vertex. (Prompt decays require a large mass splitting and are less motivated given Eq.~\eqref{eq:coannn}.) Neither of these signatures has been explored in dedicated experimental analyses so far. 

We observe that, when $\Delta m_\chi < m_{\pi^0}$, the decay $\chi_2 \to \chi_1 \gamma\gamma\gamma$ through an off-shell $\pi^0$ gives a lifetime $\gtrsim \mathcal{O}(1)$\,sec where the cosmological bounds kick in. Thus, the interesting mass range for colliders is $\Delta m_\chi > m_{\pi^0}$. We also found the decay rate for $\chi_2 \to \chi_1 \gamma$, na\"ively induced at 1-loop, to be zero.

\begin{figure}[t]
    \centering
    \includegraphics[width=0.45\textwidth]{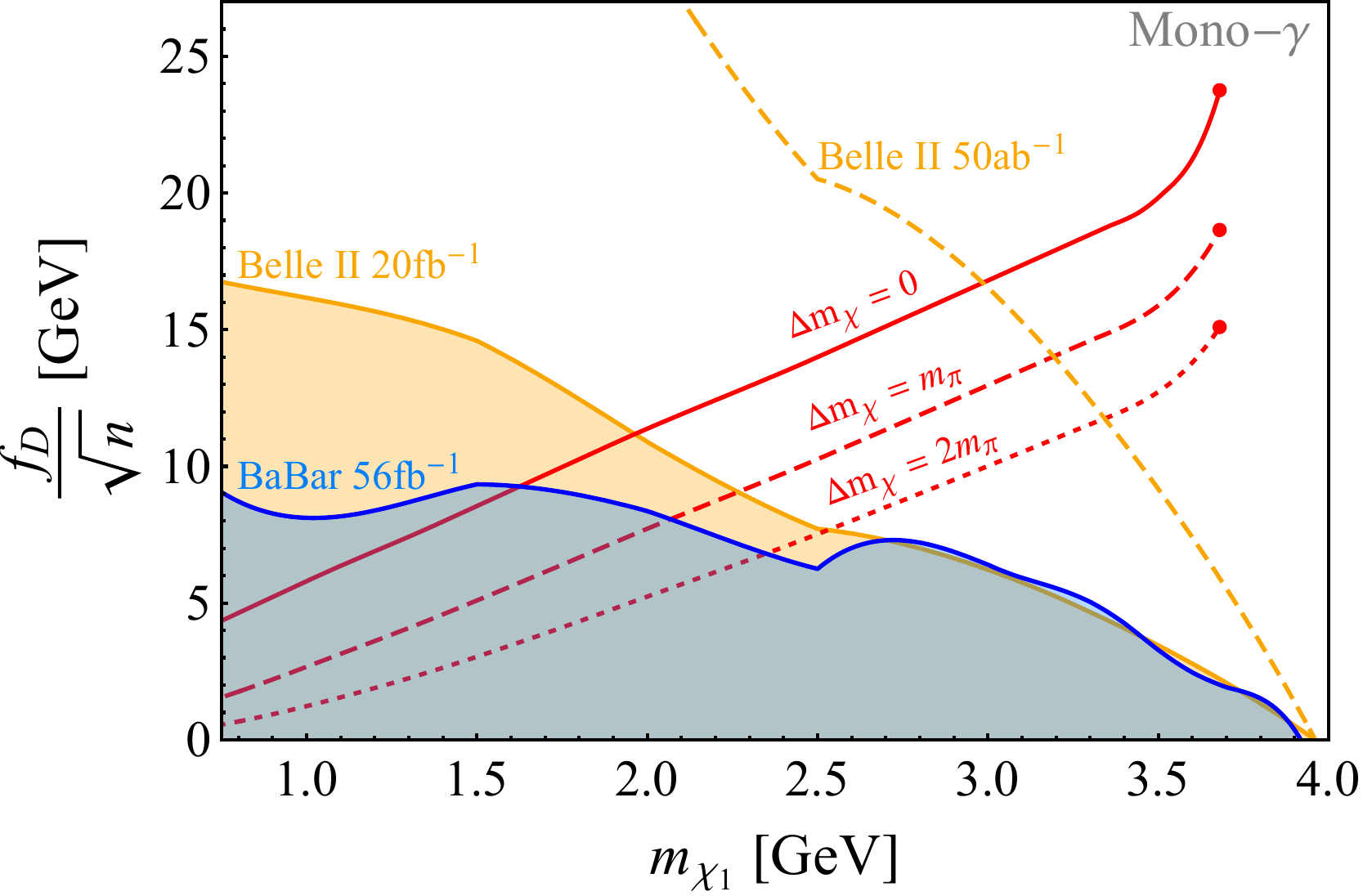}
    \caption{Limits by the B-factories on the parameter. The signature corresponds to $\pi^0+\slashed{E}_T$, where $\pi^0$ is identified as a photon such that mono-$\gamma$ searches are used. The red lines delineate the parameter space points giving the correct DM relic abundance, for different $\Delta m_{\chi}$ values. Both Belle II exclusion lines correspond to the projections~\cite{Belle-II:2018jsg}, while BaBar already performed a mono-$\gamma$ search~\cite{BaBar:2014zli}.}
    \label{fig:BelleII_BaBar}
\end{figure}

To illustrate the potential of Belle II, we focus on the small $\Delta m_\chi$ scenario (detector-stable $\chi_2$). We leave the other case for future work. The energy of $\pi^0$ originating from $e^+e^-\to\gamma^*\to\pi^0\chi_1\chi_2$ is such that the two photons from $\pi^0\to\gamma\gamma$ cannot be sufficiently separated and are detected as a single photon~\cite{Belle-II:2018jsg}. Therefore, we recast the existing (proposed) \textit{dark photon} search at BaBar~\cite{BaBar:2014zli} (Belle II~\cite{Belle-II:2018jsg}) in the \textit{mono-$\gamma$} channel, keeping in mind that a dedicated analysis would outperform our study, see also~\cite{Duerr:2019dmv}. We simulate the signal events using \aNLO~\cite{Alwall:2014hca} and apply the same event selection as reported in~\cite{Belle-II:2018jsg}. The overall analysis efficiency is slightly less compared to the dark photon search, resulting in the limits shown in Fig~\ref{fig:BelleII_BaBar}. 

The red lines in the figure produce the correct DM relic abundance for three values of the mass splitting, and all terminate at $m_{\chi_1} \lesssim 3.7$ GeV as discussed in Sec.~\ref{sec:relic}. Already with $20\,\rm{fb}^{-1}$, Belle II can test a significant portion of the parameter space, slightly more than the existing BaBar exclusion based on $56\,\rm{fb}^{-1}$. This is due to the improved single photon trigger efficiency at Belle II. The projections for the full luminosity of $50\,\rm{ab}^{-1}$ (dashed orange) are extremely promising, covering most of the parameter space of interest. Finally, should Belle II discover such a signal in $\pi^0 \chi_1\chi_2$, then observing the production of $\eta \chi_1 \chi_2$ would serve as the next target (and a `smoking gun' for this topological portal), since it is predicted to occur at a fixed rate relative to the pion channel -- see Eq.~\eqref{eq:action-new-2}.
Finally, it is important to note that the targeted $\chi_i$ mass range for Belle II exceeds the reliable scope of our EFT description of the topological portal. An analysis involving a concrete UV completion is necessary to address this limitation, which is left for future work.


\section{Outlook}

In this letter, we postulate a {\em unique} topological portal operator between QCD and a dark QCD-like sector, which successfully realizes the light thermal inelastic DM scenario while offering exciting signatures at Belle II.

Future explorations will be on two fronts. Firstly, developing an ultraviolet completion for our topological operator is crucial. We anticipate the involvement of a new state that mediates interactions between quarks and the dark sector. This should result in correlated signatures observable in other experiments, such as those conducted at the LHC. Secondly, whether Nature has selected this unique and subtle portal for primary communication between the visible and the dark sector is a question that Belle II could resolve by designing specific searches for its distinctive signatures, a course of action we highly recommend for the collaboration.

\section*{Acknowledgement}

We are grateful to Tim Cohen, Javier M. Lizana, and especially Nakarin Lohitsiri for discussions. We further thank Nakarin Lohitsiri for providing helpful comments on the manuscript. 
NS has received funding
from the INFN Iniziative Specifica APINE. The work of JD and NS was partially carried out at the University of Z\"urich, supported by the European Research Council (ERC) under the European Union’s Horizon 2020 research and innovation programme under grant agreement 833280 (FLAY) and by the Swiss National Science Foundation (SNF) under contract 200020-204428. AG acknowledges the SNF for funding through the Eccellenza Professorial Fellowship ``Flavor Physics at the High Energy Frontier'' project number 186866. 

\section*{End Matter}

Here, we elaborate on the mathematical basis of the topological portal interaction studied in this paper.

In low-energy EFTs that describe Goldstone bosons on target space $X$, a class of topological terms that generalize the original WZW term~\cite{Wess:1971yu, Witten:1983tw} can be obtained by integrating (possibly locally-defined) differential forms on $X$~\cite{Alvarez1985,DHoker1994}, equipped with some condition for $G$-invariance. Differential cohomology and invariant versions thereof provide one precise way to classify such terms~\cite{Davighi:2020vcm}, and refinements of this via bordism have been developed in recent work~\cite{Yonekura:2020upo}, but we shall not need such sophistication here.

The key is that, for a 4-d QFT describing Goldstones on $X$, a WZW-like topological term can be constructed given any integral, closed differential 5-form. Because we are interested in $G$-invariant actions where $G$ is moreover assumed to be a semi-simple group, $G$-invariance of the closed 5-form is enough to guarantee $G$-invariance of the WZW action~\cite{Davighi:2018inx,Davighi:2020vcm,DHoker1994}. We here study product cosets linking QCD to a dark sector, of the form~\eqref{eq:coset}. The $SU(3)_L\times SU(3)_R$-invariant closed 5-form $\omega_5$ determines the usual WZW term for QCD, as we recounted in the main text.

Our interest is in a topological WZW-like interaction\footnote{We ignore `theta-like' topological terms corresponding to closed 4-forms because these are locally total derivatives that cannot give rise to new local interactions with dark matter.} that can serve as a portal between the QCD and dark pions, which should therefore involve both types of field. The candidate closed 5-form should therefore admit a factorization, as the wedge product of a $k$-form $\omega_k$ on $SU(3)_L\times SU(3)_R/SU(3)_{L+R}$ with a $(5-k)$-form $\Omega_{5-k}$ on $K/H$. Because of the product structure of $X$, it is straightforward to show that closedness, integrality, and invariance properties can be separately required of $\omega_k$ and $\Omega_{5-k}$.

\begin{table}[t]
  \centering
  \begin{tabular}{|c|ccccc|}
    \hline
    & & \cellcolor{blue!30}Portal & & & \cellcolor{orange!40}SIMP \\
    $p$ & $~~1~~$ & \cellcolor{blue!30}$2$ & $~~3~~$ & $~~4~~$ & \cellcolor{orange!40}$5$ \\
    \hline
    $H^p(SU(2))$ & $0$ & $0$ & $\R$ & $-$ & $-$ \\
    $H^p(SU(n)),\, n\geq 3$ & $0$ & $0$ & $\R$ & $0$ & \cellcolor{orange!40}$\R$ \\
    \hline
    $H^p(SU(2)/SO(2))$ & $0$ & \cellcolor{blue!30}$\R$ & $-$ & $-$ & $-$ \\
    $H^p(SU(3)/SO(3))$ & $0$ & $0$ & $0$ & $0$ & \cellcolor{orange!40}$\R$ \\
    $H^p(SU(4)/SO(4))$ & $0$ & $0$ & $0$ & $\R$ & \cellcolor{orange!40}$\R$ \\
    $H^p(SU(n)/SO(n)),\, n\geq 5$ & $0$ & $0$ & $0$ & $0$ & \cellcolor{orange!40}$\R$ \\
    \hline
    $H^p(SU(2n)/Sp(2n)),\, n\geq 2$ & $0$ & $0$ & $0$ & $0$ & \cellcolor{orange!40}$\R$ \\
    \hline
  \end{tabular}
  \caption{The de Rham cohomology groups in low degrees for all homogeneous spaces that are expected to arise as pNGB manifolds in QCD-like theories~\cite{cartan1959demonstration}. Entries shaded blue (of which there is only one) feature a topological portal interaction with the pions of QCD; entries shaded orange feature a 5-point dark pion WZW term, as employed in the strongly-interacting-massive-particle (SIMP) mechanism~\cite{Hochberg:2014dra,Hochberg:2014kqa}. No coset features both portal and SIMP interactions. \label{tab:dR-results} }
\end{table}

Besides the invariant closed 5-form $\omega_5$, QCD features only one other invariant differential form, that is the closed 3-form $\omega_3 \propto \mathrm{Tr} (g^{-1} dg)^3$, as discussed in the main text.
In the pure QCD action, $\omega_3$ does not feature simply by virtue of its degree; it does, however, appear as the topologically-conserved current in QCD, which can be identified with baryon number~\cite{Witten1983}.
In our dark sector extension of QCD, the 3-form $\omega_3$ becomes the unique QCD-invariant form we have at our disposal with which to try and form a non-trivial topological portal that links QCD pions to the dark sector.
Such a mixed topological term therefore requires there be a $K$-invariant closed 2-form $\Omega_2$ on the dark coset $K/H$, which we can `wedge' with $\omega_3$ to form a $G$-invariant closed 5-form.

Which dark sector cosets $K/H$ feature such an invariant 2-form? For now, let us suppose that the $K\to H$ breaking transition is due to chiral symmetry breaking in the dark sector. Then the viable coset patterns take the form $SU(N)$, $SU(N)/SO(N)$, or $SU(2N)/Sp(2N)$. (One could, in principle, consider other options such as a complex projective space, $\mathbb{C}P^n$, that go beyond the chiral symmetry breaking dynamical assumption.)
Because these $K/H$ are not just homogeneous spaces but, moreover, symmetric spaces, one can show that invariant forms are in 1-to-1 with 
de Rham cohomology classes~\cite{Chevalley1948,schwarz1994}. Therefore, a dark coset $K/H$ features a topological portal interaction iff $H^2_{\mathrm{dR}}(K/H) \neq 0$. From Tab.~\ref{tab:dR-results}, we find there is a {\em unique choice} of dark coset within the classes we consider that features such a topological portal term, which is 
\begin{equation}
    K/H = SU(2)/SO(2) \cong S^2
    \, .\vspace{0.2cm}
\end{equation}
The closed, $K$-invariant 2-form in question is simply the unit-normalized volume form $\Omega_2$ on $S^2$, that we use in the main text to construct the topological portal as in~\eqref{eq:action-new}.



Just as is the case for the familiar WZW term of QCD, it is crucial to gauge the electromagnetic subgroup $U(1)_Q \subset SU(3)_{L+R}$, generated by $Q=t_3+t_8/\sqrt{3}$, in order to arrive at the Lagrangian~\eqref{eq:action-new} and thence derive the leading phenomenological consequences of this Lagrangian. While the gauging of WZW terms is in general complicated~\cite{Witten1983, Hull1989, HULL1991379, Garcia-Compean:2010xgv, Brauner:2018zwr}, here we wish to gauge only an abelian subgroup in an effective 2d WZW term --- the QCD part. 
The procedure here simplifies to shifting $\omega_3$ by
$\frac{-e}{4\pi^2} F \wedge \mathrm{Tr}\left(Q g^{-1}dg\right)$ where $F=dA$ is the QED field strength (see {\em e.g.}~\cite{Yonekura:2020upo}). Evaluating the trace picks up contributions only from the $\pi^0$ and $\eta$.
Wedging with $\Omega_2$ to form the gauged portal interaction 5-form, 
we obtain the term given in the second line of~\eqref{eq:action-new-2}.


\bibliographystyle{JHEP}
\bibliography{references.bib}

\end{document}